\documentclass{iopjournal}
\usepackage{amsmath}
 \usepackage{amssymb}
\usepackage{siunitx}
\usepackage{enumerate}
\usepackage[
    backend=biber, 
    natbib=true,
    style=numeric,
    sorting=none
]{biblatex}
\begin{document}

\articletype{Paper} \\ % e.g. Paper, Letter, Topical Review...
\title{Adaptable phase retrieval for coherent transition radiation
spectroscopy based on differentiable physics information}

\author{Ritz Ann Aguilar$^{1,*}$ \orcid{0000-0002-4166-9507}, 
Maxwell LaBerge$^1$ \orcid{0000-0003-3089-4087},
Andreas Doepp$^2$ \orcid{0000-0003-2913-5729}, 
Alexander Debus$^1$ \orcid{0000-0002-3844-3697}, 
Zewu Bi$^2$ \orcid{0000-0000-0000-0000}, 
Michael Bussmann$^{3,1}$ \orcid{0000-0002-8258-3881},
% Stephan Karsch$^2$\orcid{0000-0000-0000-0000}, 
Arie Irman$^1$ \orcid{0000-0002-4626-0049}, 
Ulrich Schramm$^{1,4}$ \orcid{0000-0003-0390-7671}},
Jeffrey Kelling$^{1,5}$ \orcid{0000-0003-1761-2591}

\affil{$^1$Institute of Radiation Physics, Helmholtz-Zentrum Dresden-Rossendorf, Dresden, Germany}
\affil{$^2$Centre for Advanced Laser Applications, Ludwig-Maximilians-Universität München, Garching, Germany}
\affil{$^3$Center for Advanced Systems Understanding, Görlitz, Germany}
\affil{$^4$Technische Universität Dresden, Dresden, Germany}
\affil{$^5$Technische Universität Chemnitz , Chemnitz, Germany}

\affil{$^*$Author to whom any correspondence should be addressed.}

\email{r.aguilar@hzdr.de}


\begin{abstract}
Coherent transition radiation (CTR) spectroscopy is a critical diagnostic for characterizing the longitudinal structure of relativistic electron bunches in laser-plasma and conventional accelerators. 
In practice, recovering the bunch profile from a measured CTR spectrum is an ill-posed phase-retrieval problem. 
Traditionally, this is addressed using Gerchberg–Saxton (GS)-type iterative algorithms. 
However, these implementations often rely on explicit inverse propagators, making them difficult to adapt to sophisticated experimental forward models.
In this work, we introduce a flexible gradient-based framework for CTR phase retrieval. 
By leveraging a differentiable forward model, we propose a phase-only gradient descent (GD-Phase) approach that enforces the measured spectral amplitude as a hard constraint while optimizing the Fourier phase under physical real-space priors.
%(positivity, finite support, and smoothness). 
Using synthetic CTR spectra spanning multi-peaked and strongly modulated profiles, we benchmark GD-Phase against traditional GS and a real-space amplitude-parametrized gradient descent (GD-Amp) algorithm. 
Unlike traditional methods, this formulation allows for the seamless inclusion of arbitrary differentiable experimental effects into the reconstruction loop.
We demonstrate that this physics-informed approach not only reproduces the fidelity of GS methods but also establishes a robust baseline for incorporating multi-diagnostic constraints and uncertainty quantification.
This enables the systematic extension to higher-dimensional, multimodal, and uncertainty-aware diagnostics, facilitating fast and scalable phase retrieval in realistic experimental settings.
\end{abstract}

%%%%%%%%%%%%%%%%%%%%%%%%%%%%
%% INTRO
%%%%%%%%%%%%%%%%%%%%%%%%%%%%
\section{Introduction}
Accurate measurement of the longitudinal electron beam charge distribution is essential
for understanding and controlling complex beam dynamics in modern accelerators.
High-resolution information regarding micro-structures and pre-bunching is
critical, as these features directly govern peak current and gain that are
fundamental design parameters for free-electron lasers (FELs). 
This need is particularly pressing in the nascent field of laser-plasma accelerator
(LPA)-driven FELs \cite{Wang2021, Labat2023, Pompili2022,
Barber2025}, where longitudinal beam profiles also elucidate aspects of the complex beam shaping undergone throughout this acceleration
process \cite{Lundh2013, Heigoldt2015, LaBerge2024}. 
Downstream of the accelerator, ultrashort-scale longitudinal features are shaped by
chromatic effects and transverse beam dynamics during transport, underscoring
the necessity for reliable, high-fidelity profile diagnostics \cite{Lin2012}.

Coherent transition radiation (CTR) spectroscopy has emerged as a cornerstone diagnostic
for resolving the longitudinal structure of relativistic electron bunches in
both LPAs and conventional RF-based facilities  \cite{Behrens2012,
Maxwell2013, Downer2018, Schmidt2020, Zarini2022}.  
% \cite{Behrens2012, Lundh2013, Maxwell2013, Heigoldt2015, Downer2018, Schmidt2020, Zarini2022}.  
This frequency domain workhorse excels where time domain measurements struggle. For instance, electro-optic sampling is typically restricted to a resolution of tens of femtoseconds (fs)
\cite{Wang2017}, rendering it incapable of resolving the micron-- to
sub-micron--scale structures inherent in LPA beams or the high-brightness bunches
required for modern X-ray FELs. 
While state-of-the-art transverse deflecting cavities (TDCs) can resolve single-fs features \cite{Behrens2014}, their significant footprint (spanning several meters)
limits their flexibility and makes them impractical for compact accelerator
designs.

In a typical CTR spectroscopy measurement, the detector records a spectral intensity proportional to the squared modulus of the electron bunch form factor.
Since only the magnitude is observed, the inversion becomes a phase-retrieval problem --- the spectral phase is lost, and the mapping from spectrum to the real-space density is non-unique due to inherent symmetries (e.g.\ translation and reflection) and, more generally, the ambiguities of phaseless Fourier data \cite{Fienup1982}.
A historically important approach is based on Kramers--Kronig (KK) analysis \cite{Lai1995, Lai1996}, which reconstructs a \emph{minimal-phase} spectrum by relating the spectral phase to a Hilbert-transform (principal-value) integral of the log-amplitude. 
In practice, KK requires assumptions or extrapolation outside the measured frequency range and can become unreliable for structured or multi-peaked bunch profiles where the minimal-phase assumption is violated, leading to systematic distortions even when the spectrum is otherwise well calibrated \cite{Bajlekov2013}.

To address these limitations, iterative phase-retrieval methods based on alternating enforcement of constraints in Fourier and real space \cite{Fienup1982, Gerchberg1972} have become standard in beam diagnostics.
In this family, Gerchberg--Saxton (GS) enforces the measured Fourier modulus while imposing physics-motivated constraints in the longitudinal domain. %(positivity, approximate support, smoothness).
Building on this idea, accelerator-oriented ``wrap''-type procedures augment GS-- and Hybrid Input-Output (HIO)--style iterations with additional constraints and selection strategies tailored to finite-bandwidth CTR measurements 
\cite{Heigoldt2015, Bajlekov2013, Zarini2018}.
While classical GS–type algorithms are sufficient for standalone CTR phase retrieval, they increasingly require inference workflows that extend beyond isolated reconstructions. 
Furthermore, the inverse problem remains highly non-convex, i.e., multiple distinct profiles could fit the same measured spectrum comparably well, motivating ensemble reconstructions and downstream analysis rather than relying on a single solution.

At the same time, modern accelerator facilities increasingly demand fast, robust, and automation-friendly diagnostics, including machine learning (ML)-based virtual diagnostics for real-time beam characterization and feedback \cite{Scheinker2021, Arpaia2021}.
This demand aligns with a broader shift toward physics-informed optimization and differentiable modeling in accelerator science, where diagnostic forward models and beam dynamics are embedded into differentiable programming frameworks to enable gradient-based inference \cite{Roussel2023, Kaiser2023, Huhn2025, Roussel2024, Roussel2025}.
In parallel, high-fidelity physics-based reconstructions have also been pursued using global, population-based optimization when gradients are impractical.
For example, coherent three-dimensional charge structures were reconstructed from multispectral coherent optical transition radiation images using a physics forward model together with differential evolution in \cite{LaBerge2024}.
Such approaches can be powerful, but their computational cost can become substantial when reconstructions must be repeated rapidly or used inside higher-level optimization and control loops.
In wave-based imaging more broadly, differentiable formulations have enabled efficient gradient-driven phase retrieval and physics-informed learning \cite{Wang2020, Zhang2021, Lee2023}.
Yet, analogous frameworks tailored to one-dimensional (1D) longitudinal bunch diagnostics from CTR spectra remain comparatively less explored.

In this work, we bridge this gap by implementing a physics-based forward model, within a differentiable programming framework, for 1D CTR phase retrieval.
By exploiting differentiable physics engines that use gradient-based predictive optimization as a core building block, our approach extends beyond isolated reconstructions to enable seamless composition with accurate beamline and detector models.
This approach serves a dual purpose: it provides the gradients necessary for the optimization loop and naturally supports high-throughput ensemble reconstruction via batched GPU execution.
Specifically, we introduce a phase-only gradient descent (GD-Phase) algorithm that optimizes the spectral phase as a trainable parameter while enforcing the measured spectral amplitude as a hard constraint. 
To evaluate the efficacy of this differentiable approach, we perform a rigorous benchmark against both a secondary gradient-based baseline, a real-space amplitude-parametrized gradient descent (GD-Amp), and the classical GS algorithm. 
Furthermore, we introduce a canonicalization procedure to analyze ensemble stability, providing a robust methodology for interpreting CTR reconstructions and managing inherent phase ambiguities. 
Our framework not only matches the fidelity of traditional GS methods but also provides a scalable, physics-informed path toward real-time diagnostics in modern accelerator facilities.

\begin{figure}[t]
  \centering
  \includegraphics[width=0.45\textwidth]{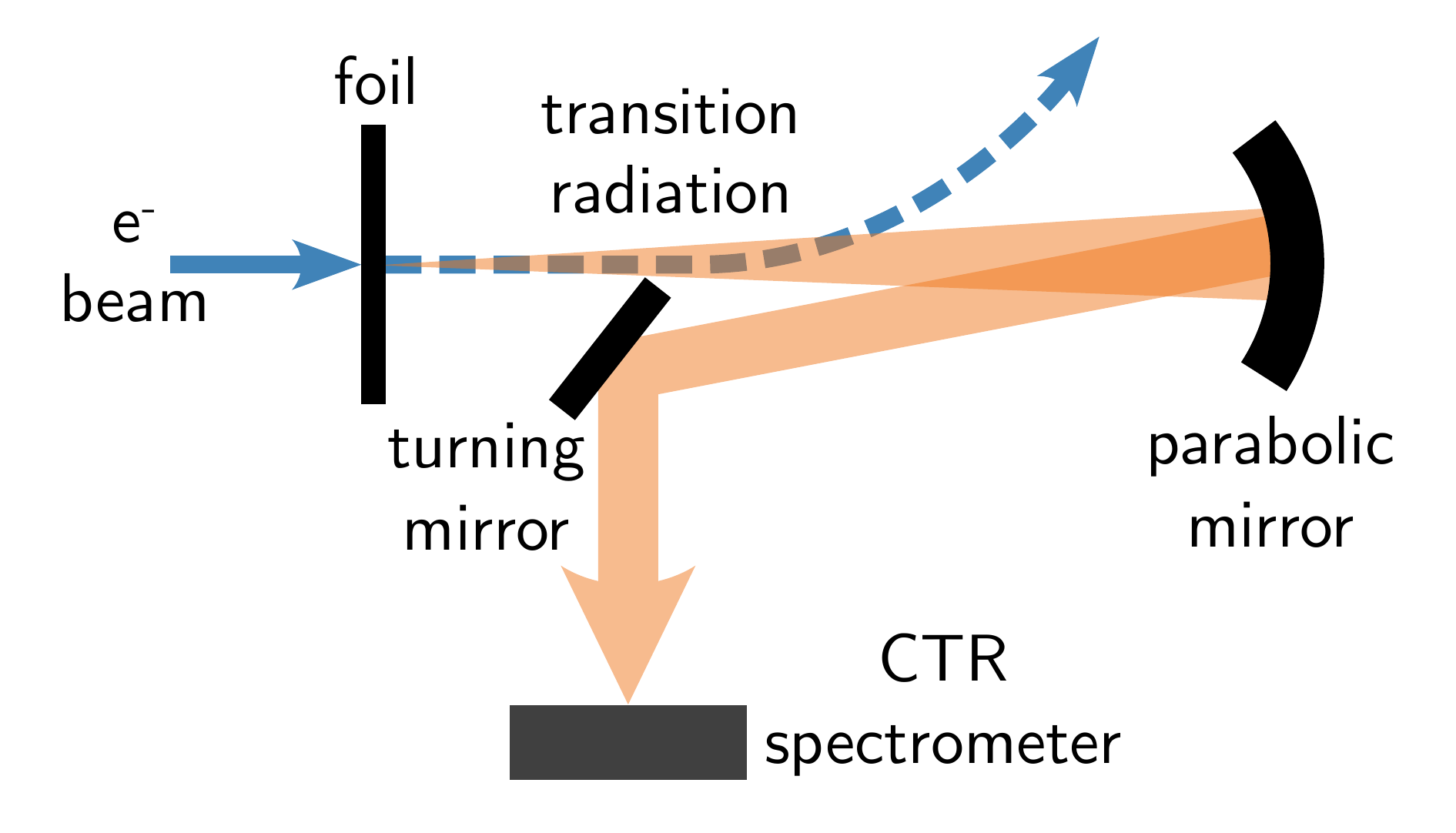}
  \caption{
  Schematic of the CTR spectroscopy setup. A relativistic electron bunch (solid blue) impinges on a metallic foil, emitting CTR into a cone (orange). A parabolic mirror collects and directs the radiation; a turning mirror steers it into the CTR spectrometer for far-field spectral measurement. The electron beam (blue dashed trajectory) is deflected by a dipole magnet and transported to an electron spectrometer (not shown).
  }
  \label{fig:ctr_far_field_setup}
\end{figure}

%%%%%%%%%%%%%%%%%%%%%%%%%%%%%%%%%%
%% METHODS
%%%%%%%%%%%%%%%%%%%%%%%%%%%%%%%%%%
\section{Methodology and Problem Formulation}
\subsection{CTR Forward Model}
The emission of coherent transition radiation (CTR) occurs as a relativistic electron bunch traverses the interface between two media with differing dielectric constants. The spectral energy radiated per unit wavenumber $k$ and solid angle $\Omega$ for a single electron is described by the Ginzburg--Frank formula \cite{Ginzburg1946}, with a corresponding $\Omega$-integrated intensity $I_{\mathrm{single}}(k)$.
For a bunch of $N_e$ electrons with $N_e \gg 1$, the coherent emission dominates, and the total energy is scaled by the 3D bunch form factor $|F(\mathbf{k})|^2$, with $\mathbf{k}$ as the wavevector. % of magnitude $k=\2 \pi /$

In typical far-field diagnostics for ultra-relativistic ($\gamma \gg 1$), well-collimated ($\text{divergence} \, \quad \, \mathrm{mrad}$) bunches, the product of the transverse bunch size $\sigma_r$ and the collection angle $\theta$ is smaller than the wavelengths of the sampled radiation ($\sin \theta \cdot \sigma_r \, \quad \, \lambda$). 
Figure~\ref{fig:ctr_far_field_setup} shows the schematic diagram of CTR spectroscopy in this far-field regime.
Under these conditions, the form factor is separable and the transverse contribution can be treated as unity \cite{Schroeder2004}. 
Hence, we can write $F(\mathbf{k}) \approx F_{\perp}(\mathbf{k}_{\perp}) F_{\parallel}(k_z) \approx F_{\parallel}(k)$, where $F_{\perp}$ and $F_{\parallel}$ are the transverse and longitudinal form factors, respectively. 
For brevity, we hereafter denote the longitudinal form factor simply as $F(k)$. 

The total coherent intensity is then $I(k) \approx I_{\mathrm{single}}(k) \cdot N_e^2 |F(k)|^2$..
In practice, the measured signal $I_{\mathrm{meas}}(k)$ is filtered by the system's spectral response $A(k)$, which incorporates the detector efficiency and a fully coherent baseline. 
The signal can then be expressed as
\begin{equation}
  I_{\mathrm{meas}}(k) = A(k) |F(k)|^2 + \eta,
  \label{eq:separated_model}
\end{equation}
where $\eta$ represents stochastic experimental noise.
Within this ultra-relativistic, low-divergence regime, the longitudinal form factor $F(k)$ can be well approximated as the Fourier transform of the normalized longitudinal charge density $\rho(z)$
\begin{equation}
  F(k) = \int_{-\infty}^{\infty} \rho(z) e^{-ikz} dz, \quad \text{subject to} \quad \int \rho(z) dz = 1, \quad \rho(z) \geq 0 ~\forall z.
  \label{eq:longitudinal_form_factor}
\end{equation}
Crucially, this Fourier relationship is what casts the bunch reconstruction as a classic phase retrieval problem. 
Because the measurement corresponds essentially to the Fourier magnitude of the charge distribution, the inversion has traditionally been treated using iterative projection algorithms, such as Gerchberg-Saxton (GS)  \cite{Gerchberg1972}, that rely on toggling between real and reciprocal space via the Fast Fourier Transform (FFT). 
We discuss this in detail in Sec.~\ref{sec:GS}.
In this work, we utilize a fully differentiable implementation of the form factor model to generate synthetic benchmarks. 

%%%%%%%%%%%%%%%%%%%%%%%%%%%%%%%%%
%%%
\subsection{The Phase Retrieval Challenge}
\label{sec:phase_retrieval_challenge}
Reconstructing the bunch profile $\rho(z)$ from the measured spectral intensity $I_{\mathrm{meas}}(k)$ is a classic 1D phase retrieval problem. Because only the spectral magnitude $|F(k)|$ is observed, the phase $\phi(k) = \arg(F(k))$ remains unconstrained by the measurement. This inversion is non-convex and fundamentally ill-posed due to several inherent ambiguities:
\begin{enumerate}
    \item \textbf{Translation.} A shift in real space, $\rho(z - z_0)$, corresponds to a linear phase ramp in frequency space, $F(k) e^{-ikz_0}$, leaving $|F(k)|$ invariant, i.e., $\rho(z - z_0) \leftrightarrow |F(k)| e^{-ikz_0}$.
    \item \textbf{Reflection (Twin Image).} The reflected profile $\rho(-z)$ possesses the conjugate spectrum $F^*(k)$, resulting in an identical intensity $|F(k)|^2$, i.e., $\rho(-z) \leftrightarrow |F(k)| e^{i\phi(-k)}$.
    \item \textbf{Scalar Ambiguity.} While a complex global phase $e^{i\theta}$ preserves intensity, the physical requirement that $\rho(z)$ be real and non-negative restricts this ambiguity to a trivial global sign, which is resolved by the positivity constraint.
\end{enumerate}

These ambiguities imply that individual reconstructions are not meaningful in isolation, motivating ensemble-based approaches and statistical analysis of solution stability.

\subsection{Synthetic Benchmark Distributions}
To evaluate our differentiable framework against traditional GS, we define four test bunches $\rho_{\text{true}}(z)$ representing reasonable experimental scenarios:
\begin{enumerate}
    \item \textbf{Blackman-Nuttall Window.} A smooth, non-Gaussian profile to test the reconstruction of edges and flat-top features \cite{Jaskula2006}.
    \item \textbf{Double-spike.} Two Gaussian peaks, testing the ability to resolve basic multi-modal distributions. The peaks are separated by $\SI{20}{\mu m}$, which is roughly a plasma wavelength in many LPA experiments. This occurs in cases where there is injection in the first and second cavities \cite{Lundh2013}, or an electron beam drives a plasma wake that is witnessed by a second beam at the back of the first cavity \cite{Heigoldt2015}.
    \item \textbf{Triple-spike.} Three overlapping Gaussian peaks representing a single bunch with moderate internal structure through spectral interference patterns \cite{Zarini2018}. It has an increased complexity to test the ordering capabilities of the beam features.
    %test the stagnation limits of GS.
    \item \textbf{Highly Modulated (Spike Train).} Representing micro-bunched beams or wakefield-modulated profiles similar to the highly modulated bunch in \cite{Zarini2018}. 
    In our case, we want to have an equivalent of the double-spike case in the spectral domain.
\end{enumerate}

\begin{figure}[t]
  \centering
  \includegraphics[width=\textwidth]{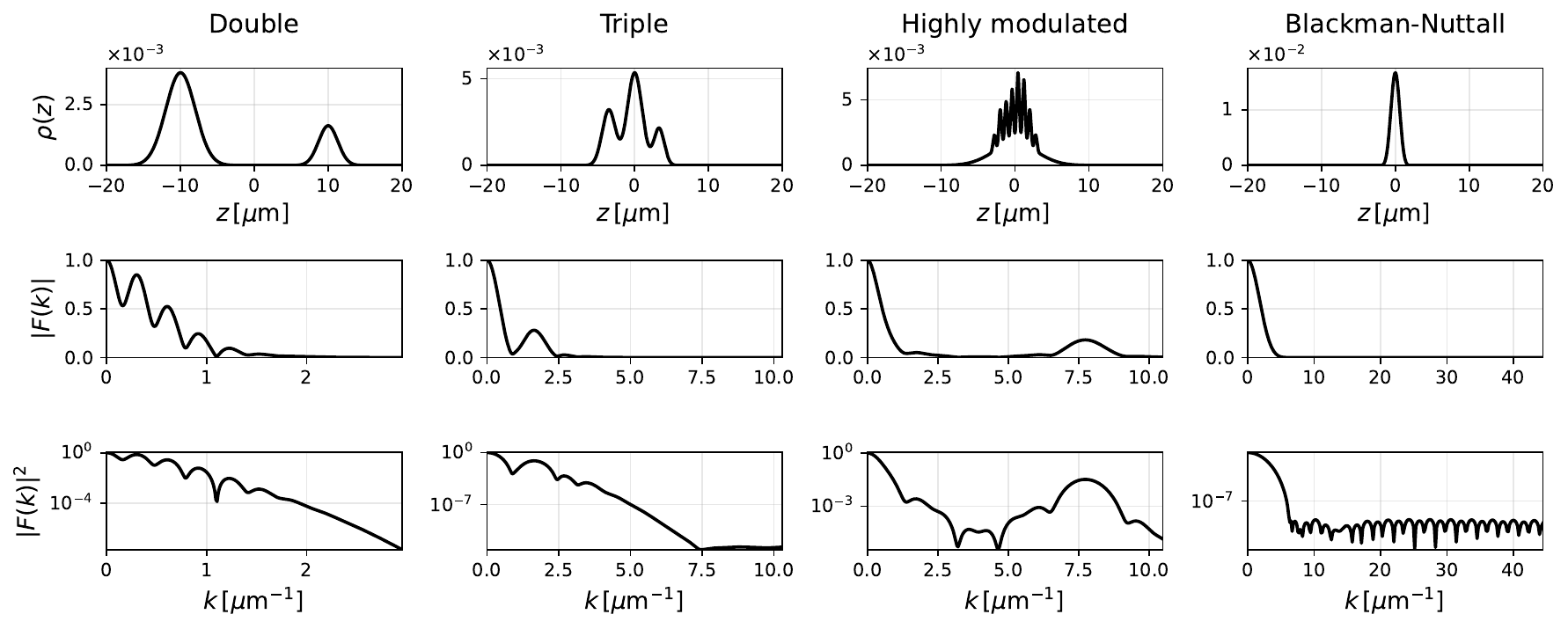}
  \caption{Synthetic benchmark suite.
  Top: representative longitudinal profiles $\rho(z)$:
  (1) double, (2) triple, (3) strongly modulated (spike train), and
  (4) Blackman--Nuttall window-like bunch, respectively.
  Middle: corresponding amplitude spectra $|F(k)|$.
  Bottom: corresponding intensity spectra $I_{\mathrm{meas}}(k)\propto |F(k)|^2$ in log-scale to emphasize the prominent features.
  The phase-retrieval task is to reconstruct $\rho(z)$ given only $I_{\mathrm{meas}}(k)$.
      % \textcolor{red}{ritz's note: add column titles; update caption}
  }
  \label{fig:bunch_families}
\end{figure}

Together, these benchmark distributions span increasing levels of non-convexity and spectral phase sensitivity, providing a controlled testbed for evaluating algorithmic conditioning and ensemble stability.
Figure~\ref{fig:bunch_families} summarizes the benchmark families and their corresponding amplitude spectra $F(k)$ and intensities in logarithmic scale $|F(k)|^2$.

%%%%%%%%%%%%%%%%%%%%%%%%%%%%%%%
%% Phase-Retrieval Algorithms
%%%%%%%%%%%%%%%%%%%%%%%%%%%%%%%
\section{Phase Retrieval Algorithms}
\label{sec:algorithms}
We compare three reconstruction strategies: the classical Gerchberg--Saxton (GS) algorithm, and two gradient-based methods: amplitude-parametrized gradient descent (GD-Amp) and phase-only gradient descent (GD-Phase). GD-Amp parametrizes the bunch directly as $\rho(z)$ and minimizes a spectral mismatch loss in terms of $|F|$ (or $|F|^2$) while GD-Phase fixes the spectral amplitude to the measured value, $|F| = \sqrt{I_{\mathrm{meas}}}$, and parametrizes only the Fourier phase, optimizing phase with respect to real-space priors
(positivity, support, smoothness, near-reality of $\rho$).
Both GD-Amp and GD-Phase are fully differentiable with respect to their parameters and can be viewed as differentiable alternatives for GS, albeit with different parametrizations.
To handle the non-convex nature of the phase retrieval landscape, all methods are executed in a multi-start setting.

\subsection{Gerchberg--Saxton (GS): The Classical Baseline}
\label{sec:GS}
The GS algorithm remains the standard for CTR inversion \cite{Heigoldt2015, Bajlekov2013, Zarini2018} precisely because it exploits the Fourier transform relationship established in Eq.~\eqref{eq:longitudinal_form_factor}. 
It functions as an alternating projection method, iteratively enforcing constraints in the spatial and spectral domains. 
In the $n$-th iteration, the update steps are:
\begin{enumerate}
    \item \textbf{Spectral Constraint:} 
    The Fourier amplitude of the current estimate is replaced by the dimensionless form factor derived from the measured spectral intensity, $|F(k)| \propto \sqrt{I_{\text{meas}}(k)}$,
    while preserving the current phase estimate $\phi(k)$. 
    \item \textbf{Spatial Constraint:} We enforce positivity $\rho(z) \geq 0$ and apply an approximate support constraint via a Tukey window $s(z)$ of half-width $z_{\max}$:
    \begin{equation}
      s(z) = \mathrm{Tukey}\!\left(\frac{z}{z_{\max}}, \alpha\right),
      \label{eq:tukey}
    \end{equation}
    with taper parameter $\alpha=0.5$, so that $\rho(z)$ decays smoothly to zero near the edges.
\end{enumerate}
While GS is computationally efficient, it exhibits several limitations in the context of modern machine learning. 
First, as an alternating projection heuristic, it relies on the existence of explicit, invertible mappings between domains, e.g., the FFT. This makes it difficult to incorporate complex, non-invertible forward models, such as detailed detector responses or neural network-based priors, into the reconstruction loop, as deriving specific projection operators for such models is often intractable.
Second, GS is prone to stagnation in local minima, particularly for profiles with disjoint support or complex modulations, as it lacks a stochastic or momentum-based mechanism to escape poor initializations. This necessitates the use of large multi-start ensembles to identify the global consensus solution.
To overcome these algorithmic constraints and enable flexible, physics-driven regularization, we turn to gradient-based optimization strategies.
We implemented a GPU version of GS in PyTorch for faster convergence to ensure a fair comparison with these gradient-based methods.

\subsection{Real-space Amplitude-parametrized Gradient Descent (GD-Amp): The Naive Differentiable Approach}
To leverage automatic differentiation, one might naively parametrize the bunch in real space, i.e.\ the profile amplitudes
$\boldsymbol{\rho} \in \mathbb{R}^N$, and minimize the spectral residual using a standard optimizer such as Adam~\cite{Kingma2014}.
Given the high dynamic range of CTR spectra, we employ a logarithmic mean-squared error (log-MSE) loss to ensure sensitivity across several orders of magnitude, particularly to high-frequency components
\begin{equation}
    \mathcal{L}_{\mathrm{Amp}}(\boldsymbol{\rho}) = \frac{1}{M} \sum_{i=1}^{M} 
    \left| 
        \log(|\mathcal{F}\{\boldsymbol{\rho}\}(k_i)| + \epsilon) - 
        \log(|F_{\mathrm{meas}}(k_i)| + \epsilon) 
    \right|^2,  
    % + \lambda \mathcal{R}(\boldsymbol{\rho}),
    \label{eq:log_mse}
\end{equation}
where $\mathcal{F}\{\boldsymbol{\rho}\}(k_i)$ is the $i$-th component of the discrete Fourier transform (DFT) of the bunch profile, $F_{\mathrm{meas}}(k_i)$ is the form factor derived from the measured intensity, and $M$ is the number of sampled spectral points.
We include a small constant $\epsilon > 0$ to ensure numerical stability within the logarithmic terms.

While intuitively appealing, this formulation is \textit{ill-conditioned}. 
The gradient of the spectral loss with respect to $\boldsymbol{\rho}$ depends nonlinearly on both the amplitude and the phase of the current estimate.
When the estimated amplitude matches the measured amplitude -- which happens quickly for the high-power, low-frequency components -- the gradient magnitude significantly diminishes.
This occurs even if the phase is entirely incorrect, causing the optimizer to stagnate. Consequently, GD-Amp often captures the gross bunch envelope but fails to resolve high-frequency micro-bunching, as will be discussed in Sec.~\ref{sec:numerical_experiments}.

\subsection{Phase-only Gradient Descent (GD-Phase): The Proposed Method}
To resolve the ill-conditioning of GD-Amp while retaining the benefits of differentiable programming (momentum, adaptivity, easy regularization), we propose the GD-Phase algorithm. This method shifts the optimization domain from the signal values $\boldsymbol{\rho}$ to the spectral phases $\boldsymbol{\phi} \in \mathbb{R}^M$. %, i.e., the phase is the primary trainable parameter.
By construction, the measured spectral amplitude is held constant, ensuring perfect data fidelity in the Fourier domain. 
The complex form factor is then defined as 
\begin{equation}
    F_{\phi}(k) = F(k) e^{i\phi(k)}.
\end{equation}

The optimization objective is no longer to fit the spectrum (which already matches by definition), but to find a set of phases $\boldsymbol{\phi}$ that produces a physically valid bunch in real space. 
The objective function $\mathcal{L}_{\mathrm{Phase}}(\phi)$ is a composite of differentiable physics-informed penalties:
\begin{equation}
  \mathcal{L}_{\mathrm{Phase}}(\phi) = \sum \lambda_{i} \mathcal{L}_{i}(\rho_{\phi}), \quad \text{with} \quad \rho_{\phi} = \mathcal{F}^{-1}\{F_{\phi}\},
\end{equation}
where $\lambda_i$ is the corresponding regularization parameter or weighting factor for each penalty term.
The individual terms are defined as follows:
\begin{itemize}
    \item \textbf{Positivity:} $\mathcal{L}_{\mathrm{pos}} = \frac{1}{N}\sum \min(0, \mathrm{Re}[\rho])^2$ penalizes unphysical negative charge.
    \item \textbf{Soft Support:} $\mathcal{L}_{\mathrm{sup}} = \frac{1}{N}\sum [(1-s(z))\mathrm{Re}[\rho]]^2$ enforces a finite bunch length via a Tukey window $s(z)$ as descibed in Eq.~\eqref{eq:tukey}.
    % \item \textbf{Smoothness:} $\mathcal{L}_{\mathrm{smooth}} = \frac{1}{N}\sum (\nabla^2 \mathrm{Re}[\rho])^2$ suppresses high-frequency artifacts.
    \item \textbf{Smoothness:} $\mathcal{L}_{\mathrm{smooth}} = \frac{1}{N}\sum ((\partial^2/\partial z^2) \mathrm{Re}[\rho])^2$ suppresses high-frequency artifacts.
    \item \textbf{Near-Reality:} $\mathcal{L}_{\mathrm{imag}} = \frac{1}{N}\sum (\mathrm{Im}[\rho])^2$ ensures the reconstructed density is a real-valued physical quantity.
\end{itemize}

This parametrization decouples the amplitude and phase. The optimizer ``surfs" the manifold of solutions that satisfy the spectral constraint, searching solely for the point on that manifold that satisfies real-space physics. This drastically reduces the dimensionality of the effective search space and avoids the vanishing gradients associated with amplitude fitting.

%%%%%%%%%%%%%%%%%%%%%%%%%%%%%%%%
%% NUMERICAL EXPERIMENTS
%%%%%%%%%%%%%%%%%%%%%%%%%%%%%%%%
\section{Numerical Experiments and Results}
\label{sec:numerical_experiments}

In the following, we analyze CTR spectroscopy phase retrieval from three complementary perspectives: (i) \emph{ensemble stability} across random initializations, (ii) \emph{conditioning and convergence behavior} under different parametrizations, and (iii) \emph{computational efficiency} for high-throughput diagnostic use cases. Rather than focusing on a single ``best'' reconstruction, we emphasize statistical robustness across multi-start ensembles as the operational metric for practical beam diagnostics. We first quantify the variability and ambiguity of the reconstructed solutions (Sec.~\ref{subsec:ensemble_analysis}), then interpret representative best-case reconstructions and convergence dynamics (Sec.~\ref{subsec:algo_performance}), and finally assess throughput and scaling (Sec.~\ref{subsec:comp_efficiency}).

\paragraph{Implementation Details.}
We denote by $r \in \{1,\dots,R\}$ the restart index and by $\rho_r(z)$ the reconstructed longitudinal charge density on a uniform grid $z_n = z_{\min} + n\Delta z$ with $n=0,\dots,N-1$, where $N$ is the number of samples and $\Delta z$ is the grid spacing. 
% The ground-truth profile is $\rho_{\mathrm{GT}}(z)$ (synthetic benchmarks).
The measured spectrum is sampled at wavenumbers $\{k_i\}_{i=1}^M$ with $M$ spectral samples, and $F(k)$ denotes the longitudinal form factor ($\mathcal{F}(\rho)$). 
All profiles are normalized such that in discretized form, $\sum_n \rho(z_n)\Delta z = 1$.

All bunch profiles are sampled on a grid of $N=4096$ samples over $L_z = 150\,\mu\text{m}$.
For the GD approaches, the optimization is performed using the Adam optimizer \cite{Kingma2014} with a learning rate of $10^{-4}$ and $10^{-2}$ for the GD-Amp and GD-Phase, respectively. 
The choices for the learning rates were determined using a simple grid search. %grid search on synthetic double-peak bunches
The GD-Phase penalty weights were selected via the same method leading to $\lambda_{\text{pos}} = \lambda_{\text{sup}} =  \lambda_{\text{imag}} = 10$, and $\lambda_{\text{smooth}} = 5$.
We note that these hyperparameters were kept constant across all benchmark distributions to evaluate the generalizability of the algorithms. 
While individual bunch morphologies (e.g., the highly modulated one) could benefit from target-specific tuning of the smoothness regularizer, a unified set of weights was found to provide a stable compromise between resolution and artifact suppression.
We observe that reconstruction quality is robust to factor-of-two changes in these hyperparameters.

All algorithms were executed with $R=32$ restarts on a single \texttt{NVIDIA V100 (32GB)} GPU to show fairness in the benchmarking analysis. 
We also employ PyTorch’s \texttt{autograd}, i.e., gradients are backpropagated through the inverse FFT to the phase parameters $\phi(k)$.
Although we could also benefit from parallelizing the runs across multiple GPUs, we leave this for future work.
For both GS and GD-Phase, the initializations across the restarts are completely random. 
The GD-Amp, on the other hand, has one of its runs initialized with the result of the GS (1 deterministic output). 
This is to show possible improvement on the GS reconstruction which has the possibility to get stuck even with increased iterations for complex distributions (without the aid of HIO algorithms). 
The rest of the restarts for the GD-Amp are then started with a fixed magnitude for the bunch profile as computed from the spectrum but with random phases (stochastic outputs).
We set the GS iterations and GD-Amp and GD-phase steps to 5000 for the benchmarking analysis.

\begin{figure}[t!]
  \centering
  \includegraphics[width=0.9\textwidth]{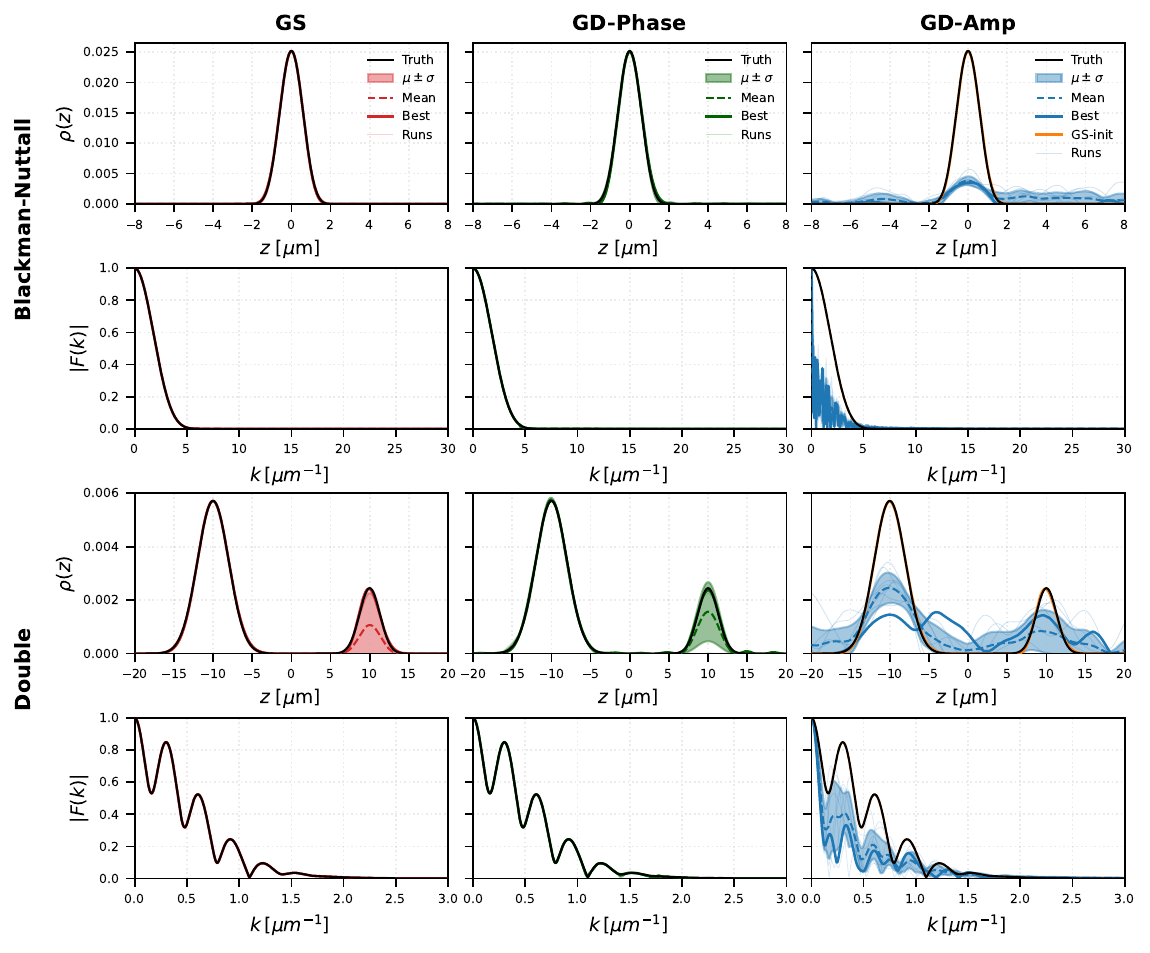}
    \caption{
        Reconstruction stability analysis across $R=32$ restarts for the relatively simple synthetic benchmark distributions (Blackman-Nuttall and Double-spike).
        Columns correspond to the three algorithms (GS, GD-Phase, GD-Amp), and rows represent the bunch profiles $\rho(z)$ and corresponding spectra $F(k)$. 
        In each panel, the true profile (\textbf{Truth}) is shown in black.
        The \textbf{Best} reconstruction (lowest spectral loss) is plotted as a solid colored line.
        For the GD-Amp, we denote by orange solid line the candidate solution initialized with the GS output for a deterministic run  (\textbf{GS-init}).
        The \textbf{Mean} of the top 30\% candidate solutions is shown as a dashed line with the shaded regions indicating the $\mu \pm \sigma$ confidence interval of these solutions, quantifying the local stability of the convergence basin.
        Thin semi-transparent lines (\textbf{Runs}) display the ensemble of the top 30\% solutions. 
        Note that for the chosen percentage of the top solutions, the spectra $F(k)$ for the GS and GD-Phase are in good agreement with the true solution; hence, no visible variance in their spectra plots.
      }
  \label{fig:gs_gd_spaghetti_a}
\end{figure}

\begin{figure}[t!]
  \centering
  \includegraphics[width=0.9\textwidth]{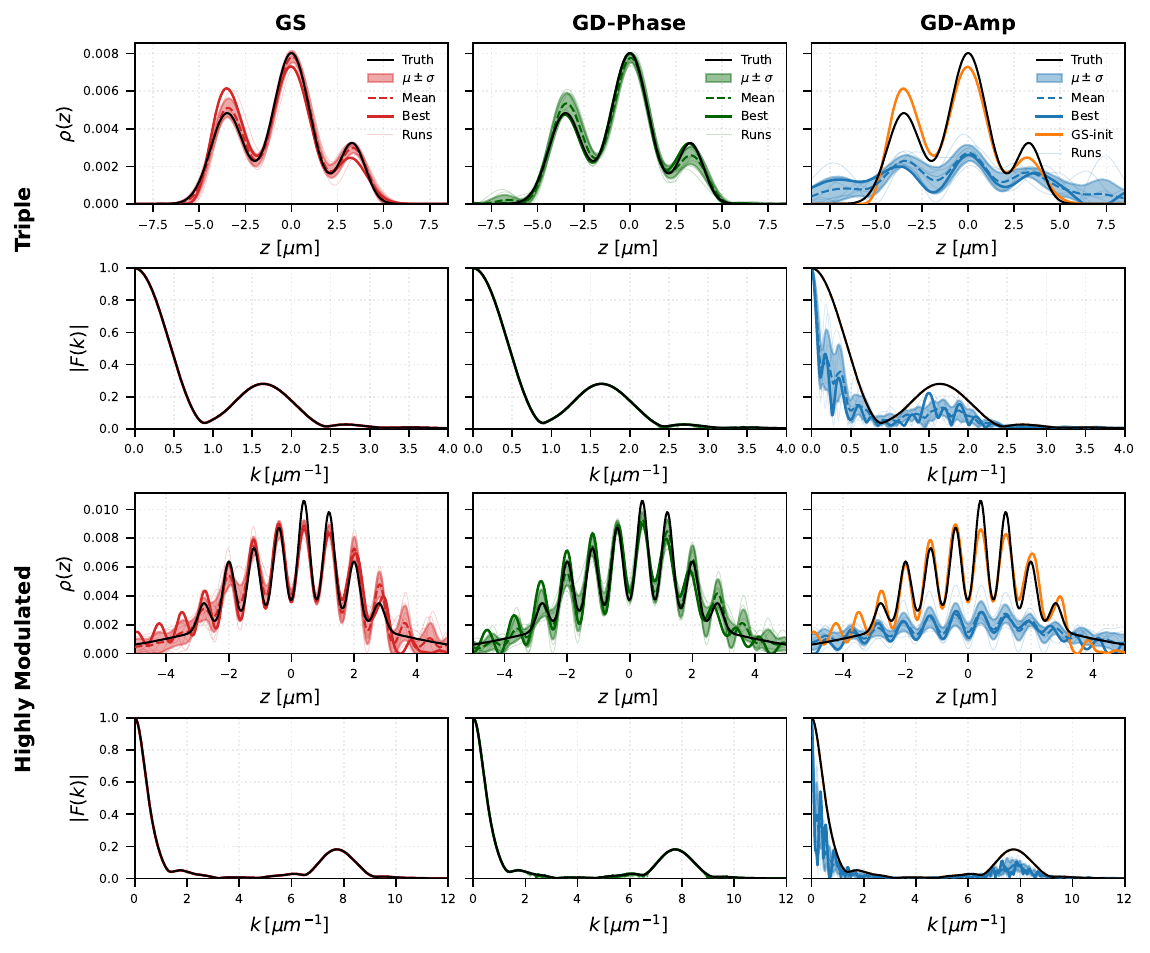}
    \caption{Reconstruction for the complex synthetic benchmark distributions (Triple-spike and Highly Modulated bunch) similar to Fig.~\ref{fig:gs_gd_spaghetti_a}. 
    Columns correspond to the three algorithms (GS, GD-Phase, GD-Amp), and rows represent the bunch profiles $\rho(z)$ and corresponding spectra $F(k)$. 
    Similar to Fig.~\ref{fig:gs_gd_spaghetti_a}, for the chosen percentage of the top solutions, the spectra $F(k)$ for the GS and GD-Phase are in good agreement with the true solution; hence, no visible variance in the spectra plots.
      }
  \label{fig:gs_gd_spaghetti_b}
\end{figure}

\paragraph{Canonicalization.}
We remove trivial degeneracies and ambiguities (detailed in Sec.~\ref{sec:phase_retrieval_challenge}) that do not affect the spectral fit. 
For each restart $r$, we align the profile $\rho_r(z)$ with the lowest-loss candidate using cross-correlation to correct for translation shifts. 
We additionally test the mirror-image $\rho_r(-z)$ and retain the orientation that maximizes spatial correlation. 
This canonicalization ensures that the remaining ensemble variance reflects genuine physical shape ambiguity rather than trivial coordinate offsets.

\paragraph{Reporting metrics.}
\label{par:reporting_metric}
Although GS is not defined as the minimizer of a single scalar objective, we evaluate all methods using 
\begin{equation}
    \mathcal{L}_{\mathrm{spec}}
=
\frac{1}{M}
\sum_{i=1}^{M}
\left[
\log\!\left(|F_r(k_i)|+\epsilon\right)
-
\log\!\left(|F_{\mathrm{meas}}(k_i)|+\epsilon\right)
\right]^2,
\label{eq:Lspec}
\end{equation}
with $\epsilon = 10^{-12}$.
Here, $F_r(k_i)$ is the form factor of the $r$-th reconstruction and $F_{\mathrm{meas}}(k_i)$ is the measured form factor magnitude (synthetic benchmarks in this work).
 Notice that this spectral loss has the same form as Eq.~\eqref{eq:log_mse}. %but without the additional physics terms.

Additonally, to quantify reconstruction quality in real space, we report the correlation coefficient between the ground truth profile $\rho_{\text{GT}}$ and the canonicalized reconstructions $\rho_r$ using
\begin{equation}
    % \mathcal{C} = \max_{\delta z} \frac{\langle \rho_{\text{rec}}(z+\delta z), \rho_{\text{GT}}(z) \rangle}{\|\rho_{\text{rec}}\| \|\rho_{\text{GT}}\|}.
    \mathcal{C} = \max_{\delta z} \frac{\langle \rho_r(z+\delta z), \rho_{\text{GT}}(z) \rangle}{\|\rho_r\| \|\rho_{\text{GT}}\|},
    \label{eq:correlation}
\end{equation}
where we use the definitions for the discrete inner product $\langle a,b\rangle = \sum_{n=0}^{N-1} a(z_n)b(z_n)\Delta z$ and norm $\|a\|=\sqrt{\langle a,a\rangle}$.
The shift $\delta z$ in Eq.~\eqref{eq:correlation} denotes a \emph{translation} applied to the reconstructed profile; in discrete form, $\delta z = m\Delta z$ with integer shift $m$.
In practice, $\max_{\delta z}$ is implemented by maximizing the cross-correlation over a bounded set of shifts $m \in [-m_{\max}, m_{\max}]$ (or equivalently via FFT-based cross-correlation). % with appropriate zero-padding and windowing to avoid wrap-around artifacts).

\subsection{Ensemble Analysis and Clustering}
\label{subsec:ensemble_analysis}

\paragraph{Ensemble stability and ambiguity.}
Phase retrieval from intensity-only spectra is inherently non-convex and non-unique, meaning that distinct initializations can converge to different local minima. 
To extract physical meaning from these stochastic reconstructions, we analyze the full ensemble of candidates rather than relying on a single deterministic output.

Figures~\ref{fig:gs_gd_spaghetti_a} and \ref{fig:gs_gd_spaghetti_b} visualize the variability of the solutions for all benchmark families and algorithms. 
In each panel, we plot the ground truth (black), the ensemble mean (dashed), and the confidence interval defined by the standard deviation of the top 30\% candidates (shaded region).
We also plot the best solutions (solid line) for each algorithm chosen by the lowest spectral loss.
For GD-Amp, we indicate the run (orange) initialized from the output of the GS algorithm.
Thin semi-transparent curves (`spaghetti plots') display top 30\% restarts, revealing both the dominant convergence basin and the presence of outliers.

This visual analysis highlights a critical conditioning gap between the methods.
As seen in Fig.~\ref{fig:gs_gd_spaghetti_a}, for the simple \textit{Blackman-Nuttall} and \textit{double-spike} case, the GS (red) and GD-Phase (green) algorithms converge reliably, indicated by tight confidence bands that hug the ground truth, while the GD-Amp (blue) has significant variance in its runs and not converging to the true solution at all.
For the \textit{triple-spike} and \textit{highly modulated} and bunches shown in Fig.~\ref{fig:gs_gd_spaghetti_b}, all algorithms exhibit a noticeable variance, often stagnating in local minima where the high-frequency modulation depth is underestimated.
In both simple and complex cases, GD-Amp seeded with the GS output barely improved the GS outcome.
On the other hand, GD-Phase (green) can maintain a remarkably tighter confidence band as seen in the profiles even for the complex bunches similar to GS (red), suggesting that the phase-only parametrization effectively smooths the local landscape, guiding the majority of restarts into the same solution basin.

Overall, the ensemble for each algorithm reveals a non-convex landscape. 
While a cluster of solutions converges toward the true profile, some restarts stagnate at clearly inferior outliers. 
This behavior confirms that both GS and GD-based methods must be treated as \textit{stochastic ensembles} rather than single deterministic outputs.

\begin{figure*}[t]
  \centering
    \includegraphics[width=\textwidth]{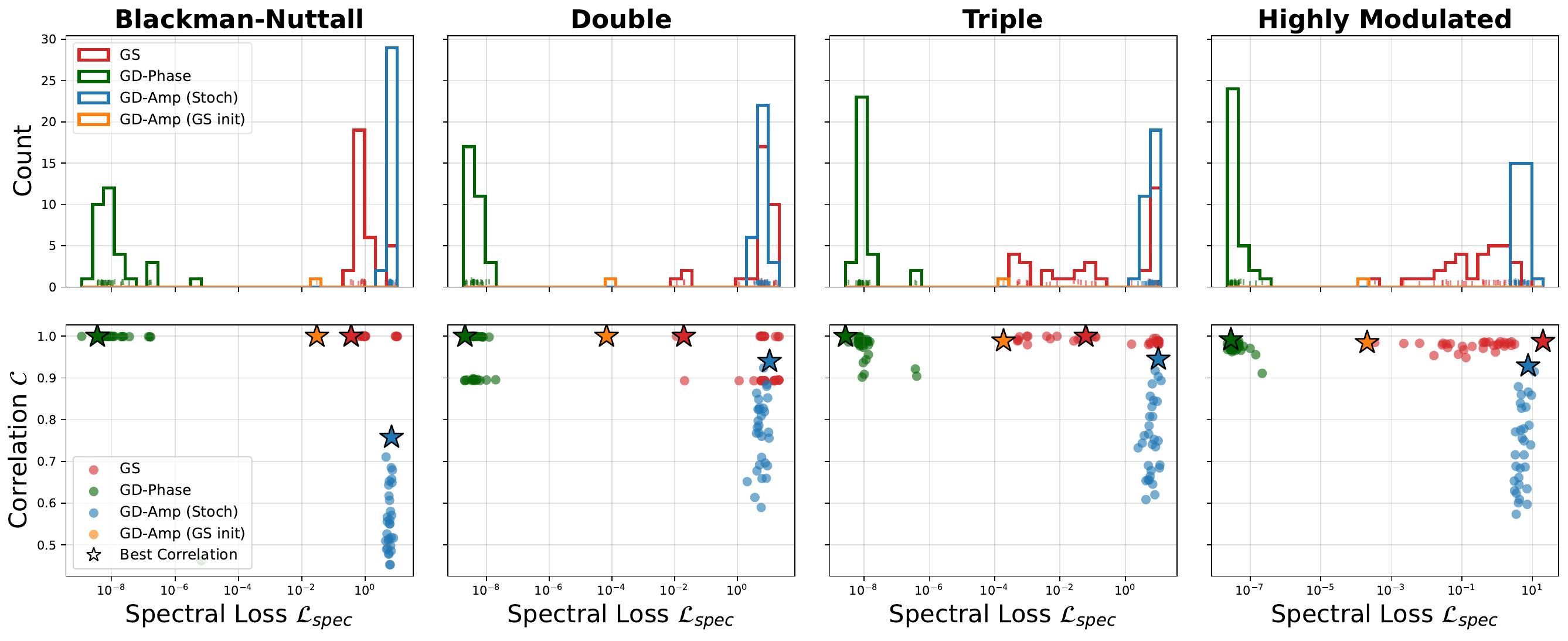}
    \caption{Statistics of the ensemble quality for all bunch families. 
      Histogram of final spectral loss $\mathcal{L}_{\text{spec}}$ over $R=32$ restarts for all algorithms (\textbf{top}).
      GD-Phase (green) consistently achieves the lowest loss mode.  GD-Amp with GS initialization (orange) shows slightly better spectral loss than GS (red).
      GD-Amp from random restarts (blue) consistently shows the highest low mode. 
      Corresponding correlation $\mathcal{C}$ versus spectral loss $\mathcal{L_\mathrm{spec}}$ (\textbf{bottom}). 
  }
  \label{fig:loss_statistics}
\end{figure*}

\paragraph{Statistical reliability.}
To quantify these variability, we analyze the distribution of spectral losses and the relationship between spectral and spatial metrics (see Sec.~\ref{par:reporting_metric}).
Figure~\ref{fig:loss_statistics} (top row) displays the histogram of $\mathcal{L}_{\text{spec}}$ over all $r$ restarts.  
The distributions are clearly multimodal, i.e., a group of low-loss solutions co-exists with higher-loss outliers.  
GD-Phase consistently has the lowest spectral loss values (clustered to the left), while GS and GD-Amp have higher loss values (clustered to the right).
These loss values may not entirely reflect what we see in Figs.~\ref{fig:gs_gd_spaghetti_a} and \ref{fig:gs_gd_spaghetti_b}, which could imply that GD-Phase, despite the larger variance across solutions, could find closer solutions to the true solution compared to GS.

We further verify the chosen spectral loss metric $\mathcal{L}_{\text{spec}}$ from Eq.~\eqref{eq:Lspec} as a valid proxy for reconstruction quality by plotting the spatial correlation coefficient $\mathcal{C}$ from Eq.~\eqref{eq:correlation} against it in Fig.~\ref{fig:loss_statistics} (bottom row). 
Notably, GD-Phase (green circles) clusters tightly in the top-left corner (high correlation, low loss), indicating a near-100\% convergence rate to the physical solution basin.
In contrast, GD-Amp followed by GS show a wider spread across correlation values, with stagnated runs for GD-Amp (e.g., $\mathcal{C} < 0.8$) corresponding to being stuck in a local minima.
Crucially, we do not observe distinct high-quality ``twin'' clusters (that would appear as low-loss, low-correlation points), confirming that our canonicalization effectively merges the time-reversed solutions.
Instead, the scattered points with low correlation correspond to stagnation points --- local minima with significantly higher spectral loss where the algorithm failed to resolve the correct bunch features.

\begin{figure}[t]
  \centering
  \includegraphics[width=\textwidth]{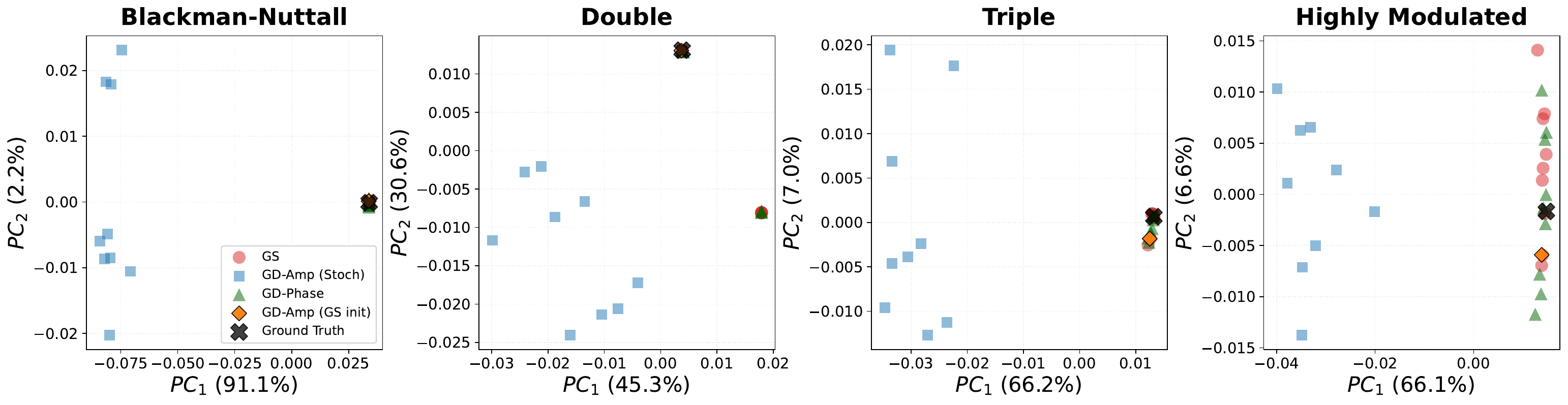}
  \caption{
    PCA landscape of reconstruction ensembles. For each bunch distribution, we project canonicalized longitudinal profiles $\rho(z)$ onto the first two principal components (PC1–PC2) of a PCA model fit to the union of the ground truth profile (black cross), the top lowest-loss restarts from each algorithm, and the seeded GD-Amp run initialized from the best GS solution (orange diamond). Clouds correspond to the top 30\% subset for GS (circles), GD-Phase (triangles), and GD-Amp with random initialization (squares). 
    The explained variance ratios shown on the axes quantify how much of the ensemble variability is captured by PC1–PC2.
  }
  \label{fig:gs_gd_pca}
\end{figure}

%-------------
\paragraph{PCA as a failure mode visualizer.}
To visualize the geometry of these failure modes, we use Principal Component Analysis (PCA) to project the canonicalized profiles onto their first two principal components (PC1, PC2).
PCA is used purely as a \emph{visualization tool}: if two reconstructions lie close together in the (PC1, PC2) plane, their underlying profiles are close in Euclidean ($\ell_2$) distance up to the information retained by the first two PCs; conversely, separated clusters correspond to distinct basins with qualitatively different morphology (e.g.\ incorrect peak separations or underestimated modulation depth).
We emphasize that PCA does not establish uniqueness or convexity; rather, it provides an interpretable low-dimensional summary that complements the distributional metrics.

Fig.~\ref{fig:gs_gd_pca} shows the PCA grid of all algorithms combined.
The axes are annotated with the explained-variance ratio, which quantifies the percentage of the total ensemble variability captured by each component.
For the simple \textit{Blackman-Nuttall} bunch, PC1 alone captures nearly $90\%$ of the variance, and the solutions from GS and GD-Phase cluster tightly on the ground truth (black cross).
The GS (red circles) and GD-Phase (green triangles) solutions form a compact cluster, mostly overlapping the ground truth as the problem is relatively convex or ``easy".
However, GD-Amp from random initialization (blue squares) shows significant scatter to the left. GD-Amp initialized from the best GS solution (orange diamond) almost overlap with the best GS solution as expected.
In contrast, for the \textit{double-spike} bunch, the variance is split more evenly (PC1 $\approx$ 45\%, PC2 $\approx$ 31\%), indicating a higher-dimensional error space where algorithms can fail in multiple distinct ways (e.g., peak separation error vs. amplitude ratio error).
GD-Amp generally exhibits large variance extending far from the true solution.
The scattered solutions correspond to the distorted local minima observed in the profile plots.
For the \textit{highly modulated} bunch, the vertical spread suggests that while most algorithms capture the envelope, they struggle to agree on the precise modulation depth.
This could demonstrate that the instability of GD-Amp is intrinsic to the parametrization, persisting even for simple, convex shapes where GS and GD-Phase succeed trivially.
Further analysis could be done with the clustering of the GS and GD-Phase solutions but this is left for future work.

% \\[12pt]
Taken together, these ensemble statistics motivate examining representative best-case reconstructions, spectral-phase diagnostics, and convergence dynamics to interpret how parametrization influences conditioning (Sec.~\ref{subsec:algo_performance}).

%%%%%%%%%%%%%%%%%
\subsection{Algorithmic Performance and Conditioning}
\label{subsec:algo_performance}
\begin{figure*}[t]
  \centering
  \includegraphics[width=\textwidth]{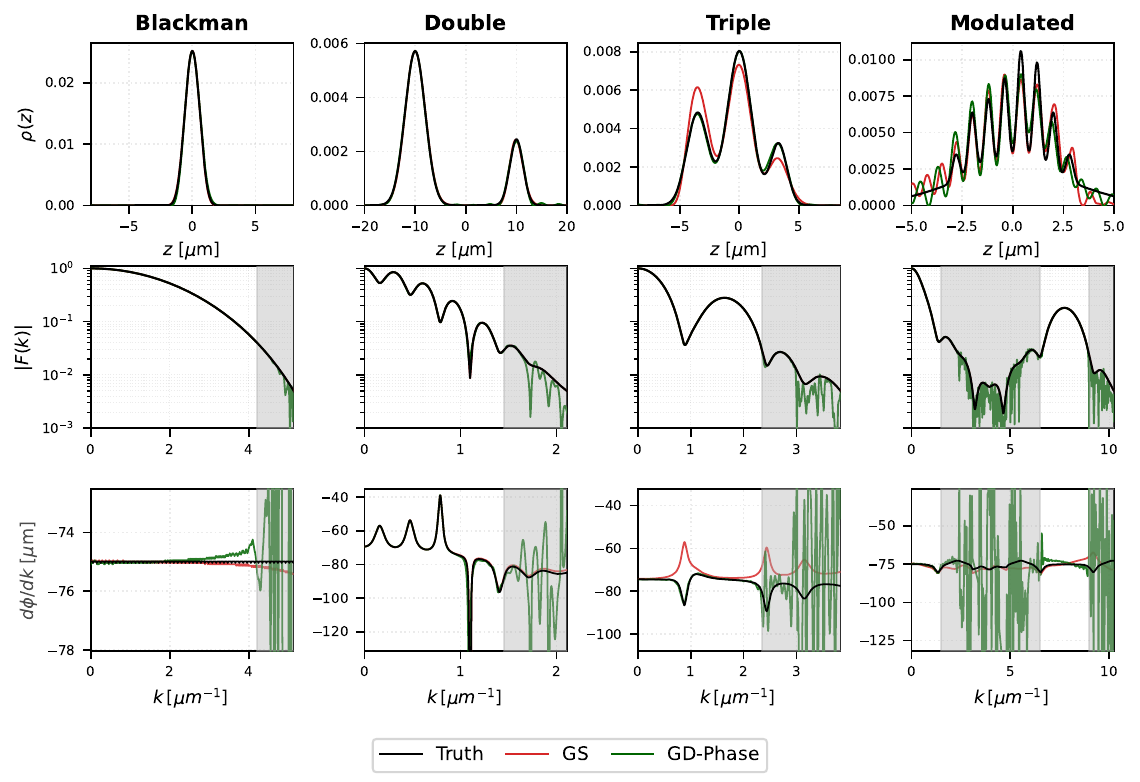}
  \caption{
    Summary of the best reconstructions (lowest spectral loss) of all longitudinal bunch profiles (\textbf{top}), their corresponding normalized spectra in logarithmic scale (\textbf{middle}), and their phase gradients (\textbf{bottom}) for the GS and GD-Phase algorithms.
    The black curve shows the ground truth while the red and green curves represent the GS and GD-Phase plots, respectively.
    The shaded regions indicate $k$-intervals where the normalized spectrum is near a floor and locally flat (in non-logarithmic scale).
    GD-Amp is omitted for better clarity of the plots.
  }
  \label{fig:overlay_phase_gradient}
\end{figure*}

\paragraph{Best-case reconstructions.} 
Figure~\ref{fig:overlay_phase_gradient} shows, for each bunch family, the best candidate (selected by lowest $\mathcal{L}_{\mathrm{spec}}$) from $R$ restarts for GS and GD-Phase, together with the normalized spectrum magnitude $|F(k)|$ and the phase-gradient $\partial\phi/\partial k$.
Note that we omit GD-Amp here for clarity because, in our experiments, its best-case solutions were substantially more variable and often visually chaotic under the same restart budget. 
Nevertheless, GD-Amp remains useful as a baseline demonstrating the ill-conditioning that arises when optimizing $\rho(z)$ directly; we report its ensemble statistics and representative outcomes separately in the previous section to avoid obscuring the clearer GS vs.\ GD-Phase comparison shown here.

For the \textit{double-spike} and \textit{Blackman--Nuttall} cases, both GS and GD-Phase recover the main peak structure and support with high fidelity (top row of Fig.~\ref{fig:overlay_phase_gradient}).
For the more challenging \textit{triple-spike} and \textit{highly modulated} profiles, the inversion becomes more ill-posed: fine oscillatory structure is difficult to recover from phaseless spectral measurements even under strong priors.
In these cases, GD-Phase more consistently preserves key morphological features (e.g.\ peak separation and envelope) compared to GS under the same restart budget, although neither method perfectly reproduces all high-frequency oscillations in the strongly modulated profile.

\paragraph{Spectral magnitude and phase gradients.}
In the spectral domain (middle row of Fig.~\ref{fig:overlay_phase_gradient}), GS reproduces the measured magnitude very accurately over the informative bandwidth.
However, magnitude agreement alone does not guarantee a correct real-space profile: multiple distinct $\rho(z)$ can yield near-identical $|F(k)|$ within finite bandwidth and in the presence of noise floors.
To probe what the measurement actually constrains beyond magnitude, we also visualize the phase-gradient $\partial\phi/\partial k$ (bottom row), computed only where the spectrum magnitude exceeds a threshold.
Across all bunch types, GS and GD-Phase agree well with the truth in the low-to-mid $k$ range where $|F(k)|$ remains appreciable, but they can diverge strongly in bands where $|F(k)|$ becomes small.

The shaded regions in Fig.~\ref{fig:overlay_phase_gradient} indicate $k$-intervals where the normalized spectrum is near a floor and locally flat, so the data provide weak additional constraints on the solution.
In these regimes, phase estimates (and especially $\partial\phi/\partial k$) are intrinsically unstable --- when $|F(k)|$ is small, tiny perturbations in complex phase can produce large apparent variations in $\partial\phi/\partial k$ without materially affecting $|F(k)|$.
Accordingly, the shaded bands are intended as visual guides for \emph{low-informativity} rather than claims of zero sensitivity; differences between algorithms in these bands should not be over-weighted in judging reconstruction quality.

\paragraph{GD-Phase conditioning.}
GD-Phase enforces the measured spectral amplitude at every step and optimizes only the spectral phase $\phi(k)$, while real-space penalties (positivity, support, smoothness, near-reality) restrict the solution to physically plausible profiles.
% This reduces the effective search space and typically concentrates multi-start solutions into fewer dominant basins, which is reflected by the more consistent morphology across bunch types in Fig.~\ref{fig:overlay_phase_gradient}.
This reduces the effective search space and, across repeated runs with different random initializations, causes the reconstructions to converge into a small number of recurring local minima. 
This is reflected by the more consistent reconstructed morphology across bunch types in Fig.~\ref{fig:overlay_phase_gradient}. 
Noticeably for the \textit{triple-spike} bunch, GD-Phase outperforms GS when these physics-motivated priors are enforced.
% Noticeably for the \textit{tripe-spike} bunch, GD-Phase outperforms GS given the physics-motivated priors. 
At the same time, the phase-gradient plots emphasize a key limitation: 
% shared by all phase-retrieval methods: 
once $|F(k)|$ approaches the measurement floor, the phase becomes underconstrained, so stability must be interpreted in the context of the informative bandwidth.
This can however be circumvented by better conditioning of the penalties and even the learning rate of the chosen optimizer, i.e., hyperparametrization of the algorithm parameters.

\begin{figure*}[t]
  \centering
    \includegraphics[width=\textwidth]{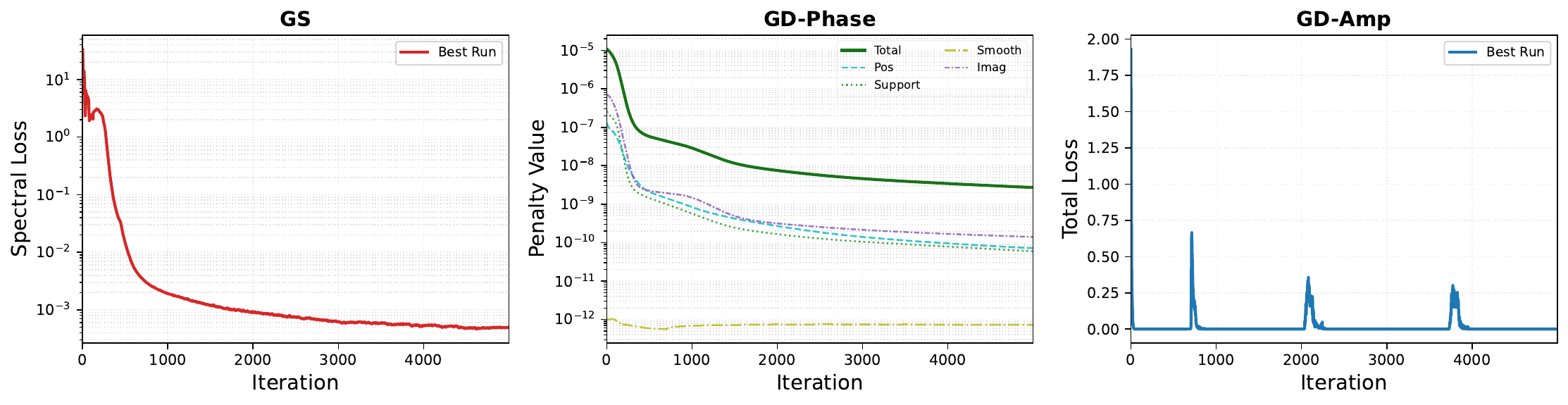}
  \caption{
    Convergence dynamics for the \textit{triple-spike} bunch. 
      \textbf{Left:} Spectral loss vs. iteration for GS.
      \textbf{Center:} Breakdown of the penalty components (Positivity, Support, Smoothness, Imaginary) for GD-Phase. Note that GD-Phase prioritizes satisfying the positivity and support constraints early in the optimization.
      \textbf{Right:} Total loss vs. iteration for GD-Amp, highlighting instability spikes characteristic of the amplitude-based parametrization with random initialization.
  }
  \label{fig:loss_dynamics}
\end{figure*}

\paragraph{Convergence dynamics.}
% Figure~6 compares typical loss traces. 
We further investigate the convergence behavior by plotting the loss traces for the best run of each algorithm in Fig.~\ref{fig:loss_dynamics}.
GS can enter stagnation-like behavior dependent on the initial phase due to its alternating-projection nature. 
GD-Amp exhibits intermittent spikes and plateaus, consistent with unstable updates in the density-parametrized landscape. 
In contrast, GD-Phase typically decreases the composite penalty more smoothly: positivity and support violations drop rapidly early in the run (guiding iterates toward a physically plausible manifold), while the smoothness term suppresses high-frequency artifacts. 
The imaginary-part penalty remains small and decays as near-reality is enforced. 
These convergence traces support the central finding of this study: \emph{parametrization strongly affects conditioning}, and phase-only optimization yields a more favorable landscape for physics-regularized CTR phase retrieval.

Overall, we see that treating the spectral amplitude as a hard constraint and optimizing only the phase $\phi(k)$, \textit{GD-Phase ensures the data constraint is satisfied exactly at every iteration.}
This effectively reduces the dimensionality of the search space and focuses the optimization entirely on satisfying real-space physical priors.

\subsection{Computational Efficiency and Scalability}
\label{subsec:comp_efficiency}
Another motivation for the proposed framework is computational throughput, particularly for enabling ensemble-based virtual diagnostics at high-repetition-rate facilities.
Table~\ref{tab:metrics_summary} summarizes the mean spectral loss $\langle \mathcal{L}_{\mathrm{spec}}\rangle$, the real-space correlation $\mathcal{C}$, and the mean wall-clock time per restart for the three algorithms (measured for $N=4096$ and $R=32$ restarts on the same hardware, using the same iteration budget) for the \textit{triple-spike} bunch.

The GS implementation is the fastest per restart because it relies on FFT-based projections and does not require backpropagation. This makes it attractive when latency is the dominant constraint and when modest ensemble sizes are sufficient.
GD-Phase incurs additional overhead from gradient backpropagation through the inverse FFT and the composite penalty terms, but remains substantially faster than GD-Amp while achieving the most consistent reconstructions under the same restart budget.

We note here that wall-clock time alone does not determine practical diagnostic cost.
The relevant quantity is \emph{time-to-reliable-solution} under stochastic multi-start which could be done in future analysis.
The ensemble statistics show that GD-Phase concentrates solutions into a dominant basin with low dispersion, whereas GD-Amp exhibits broad variability and frequent stagnation, which effectively increases the number of restarts (or tuning effort) needed to obtain a trustworthy reconstruction.
Consequently, although GS has the smallest time per restart, GD-Phase offers a favorable trade-off for scalable, differentiable ensemble reconstructions --- it preserves GPU-friendly batching and differentiability while improving robustness across random initializations.

Finally, the batched implementation enables near-linear scaling in the number of restarts $R$ on GPUs until memory limits are reached, making large ensembles practical for online monitoring and for downstream uncertainty-aware analysis.

\begin{table}[ht]
\centering
\caption{
    Statistical performance metrics ($N=4096$, $R=32$ restarts) for the \textit{triple-spike} bunch showing
    % $\sigma_{\mathcal{L}}$ denotes the standard deviation of the final loss, quantifying convergence stability. 
    % $P_{\text{succ}}$ is the probability of achieving a successful reconstruction (correlation $>0.99$).
    mean spectral loss \textbf{$\langle \mathcal{L_\mathrm{spec}} \rangle$}, correlation $\mathcal{C}$, and Time/Restart (s) representing the mean wall-clock duration per restart for all algorithms.
    % \textcolor{red}{ritz's note: update?}
}
\label{tab:metrics_summary}
\begin{tabular}{lcccc}
\hline
\textbf{Algorithm} & \textbf{Parametrization} & \textbf{$\langle \mathcal{L_\mathrm{spec}} \rangle$} & $\mathcal{C}$ & \textbf{Time/Restart (s)} \\ \hline
GS & FFT & $3.93 \pm 0.8$ & 0.988 & \textbf{0.053} \\
GD-Amp & $\rho(z)$ & $6.75 \pm 0.5$ & 0.988 & 0.787 \\
GD-Phase & $\phi(k)$ & $\mathbf{3.27 \times 10^{-8}} \pm \mathbf{2.00 \times 10^{-8}}$ & \textbf{1.000} & 0.272 \\ \hline
\end{tabular}
\end{table}

%%%%%%%%%%%%%%%%%%%%%%%%%%
%% DISC
%%%%%%%%%%%%%%%%%%%%%%%%%%
\section{Discussion}

\subsection{Comparison of Reconstruction Paradigms}
Our results highlight a fundamental trade-off between algorithmic simplicity and robustness in phase retrieval.
The classical GS algorithm remains a powerful baseline; it is simple to implement, inexpensive per iteration, and can provide high-quality solutions when the initial phase guess is fortunate.
However, it lacks a unified scalar objective function, making it difficult to integrate into differentiable machine learning pipelines or to regularize with complex physical priors. 
As shown in the \textit{triple-spike} and \textit{highly modulated} example, GS is prone to stagnation cycles where the alternating projections fail to resolve fine temporal features.

Furthermore, there is a conditioning gap between density-based and phase-based optimization.
Although GD-Amp is formally differentiable, Table~\ref{tab:metrics_summary} reveals it to be statistically unstable ($\sigma_{\mathcal{L}}=5.80$). 
This confirms that the loss landscape with respect to $\rho(z)$ is effectively flat in high-frequency directions, allowing small perturbations to cause massive divergences in the optimization path.
GD-Phase addresses these limitations and bridges this gap by reformulating the inversion as a direct minimization over the spectral phase.
By enforcing the measured spectral amplitude $F_\phi(k)=\sqrt{I_{\mathrm{meas}}(k)}\,e^{i\phi(k)}$ by construction, GD-Phase reduces the ill-posed inversion to a simpler problem: finding a phase field $\phi(k)$ that maps to a physically plausible (positive, supported, smooth) density $\rho(z)$.
This parametrization is effectively a differentiable analogue of ``alternating constraint enforcement,'' but with the distinct advantage that soft penalties (e.g., sparsity or continuity) can be added trivially to the loss function.
Numerical experiments confirm that this formulation is significantly better conditioned than optimizing the density $\rho(z)$ directly (GD-Amp).

\subsection{Relation to Advanced Reconstruction Pipelines}
Our GS implementation is intentionally minimal: it combines modulus enforcement in Fourier space with static real-space constraints (positivity, fixed tapered support, and mild smoothing).
This design isolates the core algorithmic behavior needed for a controlled comparison with the gradient-based formulation.
However, the specific challenges of CTR reconstruction such as unknown support and stagnation in local minima have motivated more sophisticated iterative schemes.
Bajlekov \textit{et al.} introduced the Bubblewrap algorithm, which solves the support problem adaptively by dynamically shrinking a constraint window around the evolving signal \cite{Bajlekov2013}.
Building on this, Zarini \textit{et al.} proposed the Foldwrap strategy for strongly modulated bunches \cite{Zarini2018}.
Their approach augments classical phase retrieval with HIO iterations to escape stagnation, adaptive support updates, and physically motivated spectral constraints (e.g., DC normalization).
Crucially, Foldwrap employs a Kramers--Kronig (KK) reference solution to post-select the most physically consistent reconstruction from a multi-start ensemble.

In this context, GD-Phase offers a differentiable alternative to these procedural enhancements.
Where Bubblewrap relies on heuristic support update rules, GD-Phase encourages compact solutions via soft penalties (e.g., support or sparsity terms) that are minimized directly by the optimizer.
Similarly, where Foldwrap uses HIO cycles to traverse the solution landscape, GD-Phase leverages the momentum and adaptive learning rates of modern optimizers (e.g., Adam) to overcome energy barriers.
While we do not incorporate KK-guided selection in this study, the differentiable framework is compatible with such physics-informed post-processing; the gradient-based priors essentially act as a soft, continuous counterpart to the hard algorithmic steps found in the state-of-the-art iterative solvers.

\subsection{Ambiguity and Multi-Diagnostic Fusion}
Despite the improved robustness of GD-Phase, the inverse problem remains fundamentally non-unique because CTR spectroscopy measures only the spectral magnitude $|F(k)|$ over a finite bandwidth and in the presence of noise floors.
In this work, we explicitly remove the trivial degeneracies via our canonicalization procedure.
As a result, the residual ensemble spread observed in the reconstruction plots and in the loss--correlation statistics reflects only ambiguity in profile morphology rather than coordinate artifacts.

Our observations motivate two complementary directions.
First, multi-diagnostic fusion can break degeneracies that are irreducible from intensity-only spectra.
Within the same differentiable framework, additional measurements can be incorporated as extra forward-model terms and losses, for example, complementary coherent-radiation spectra from different collection geometries, independent time-domain constraints, or auxiliary beamline/transport constraints that restrict physically plausible longitudinal structures.
Because GD-Phase only requires differentiability (not an analytically invertible propagator), such detector- and beamline-level fusion can be incorporated naturally in an end-to-end objective which then leads to our outlook in the next subsection.

Secondly, the framework suggests principled ways to improve GD-Amp rather than treating it as a purely negative baseline.
A key issue is that optimizing $\rho(z)$ directly couples amplitude and phase updates in $F(k)$, leading to poorly scaled gradients when the spectrum has large dynamic range.
One remedy is a \emph{hybrid constrained formulation} that combines the strengths of GD-Phase and GD-Amp:
(i) use GD-Phase to rapidly find a physically plausible phase on the hard-amplitude manifold $|F(k)| \propto \sqrt{I_{\mathrm{meas}}(k)}$, then
(ii) optionally fine-tune in real space with a density-parameterized step that preserves positivity/support while allowing limited flexibility (e.g.\ via a soft amplitude-consistency penalty restricted to informative $k$).
Such hybrids retain differentiability and may recover some of GS's practical advantages (fast modulus enforcement) while mitigating the instability observed in naive GD-Amp.

Overall, after canonicalization, the remaining uncertainty should be interpreted as an inherent ambiguity set under finite-bandwidth phaseless data.
Rather than suppressing this ambiguity, the differentiable ensemble formulation makes it explicit, enabling uncertainty-aware inference and providing a natural scaffold for incorporating additional diagnostics that render the reconstruction effectively identifiable.

\subsection{Outlook: Towards Detector-Level Inversion}
In this work, we isolated the phase-retrieval core by operating directly on the longitudinal form factor.  
However, experimental reality involves complex transport lines, diffraction effects, and detector responses that distort the signal before it is recorded. 
A key advantage of the gradient-based framework is its extensibility. 
Conceptually, the simple Fourier transform in our loop can be replaced by a fully differentiable forward model that represents the entire diagnostic beamline as a reconstruction pipeline
\begin{equation}
    \rho(z) \xrightarrow{\text{emission}} E(k, r) \xrightarrow{\text{transport}} E_{\mathrm{det}}(k, x) \xrightarrow{\text{detection}} I_{\mathrm{pred}}(k_i, x_j),
\end{equation}
where $E(k, r)$ is the radiated electric field at the source (with wavenumber $k$ and radial position $r$), $E_{\mathrm{det}}(k, x)$ is the field propagated to the detector plane at position $x$, and $I_{\mathrm{pred}}$ is the predicted intensity. The indices $(k_i, x_j)$ represent the discrete sampling of the spectral and spatial domains, respectively, as determined by the physical pixels and resolution of the detector grid.
This pipeline could simulate the CTR emission, beamline optics, and the detector response.

Unlike GS, which requires an invertible propagator (e.g., a standard FFT), GD-Phase only requires the forward model to be differentiable. This allows for the direct minimization of detector-level residuals, naturally incorporating calibration artifacts, spectral cutoffs, and instrument uncertainties into the inversion process. By treating the forward model as a computational graph, one can propagate experimental uncertainties from the raw images back to the reconstructed profile, providing a path toward rigorous error bars on real-world data.
Furthermore, while this work focuses on direct gradient-based optimization of the density profile, the differentiable framework allows for future extensions where $\rho(z)$ is parameterized by a generative model or a neural network. This would implicitly enforce structural regularity without the need for manual tuning of smoothing penalties, providing a more robust path toward automated, physics-informed beam diagnostics.

Looking ahead, the present work does not aim to replace existing CTR inversion methods in isolation, but rather to establish a robust, differentiable baseline that enables future multimodal and uncertainty-aware CTR-based diagnostics, with natural extensions toward reconstructing higher-dimensional beam distributions, including three-dimensional charge densities and ultimately six-dimensional phase spaces.

\section{Conclusions}
In this work, we introduced a differentiable, PyTorch-based framework for the reconstruction of relativistic electron bunch profiles from coherent radiation spectra. By reformulating the 1D phase-retrieval problem as a phase-only gradient descent (GD-Phase), we demonstrated a method that enforces spectral data constraints exactly while leveraging differentiable physics priors for regularization. 
Our benchmarks across diverse bunch morphologies show that GD-Phase provides a robust and scalable alternative to the traditional GS algorithm. Through the implementation of a batched GPU-accelerated pipeline, we achieved reconstruction throughputs in less than a second. Furthermore, we introduced a workflow that allows for the systematic interpretation of reconstruction ensembles, successfully navigating the inherent ambiguities of 1D phase retrieval.

This framework represents a critical step toward real-time, end-to-end differentiable virtual diagnostics for high-repetition-rate accelerators. 
By casting the inversion as gradient-based optimization through a differentiable forward model, it enables the seamless integration of arbitrary and multiple beam diagnostics together with beamline and detector models. %, whose parameters can be jointly optimized as part of the same computational graph. 
In particular, the GD-Phase parametrization stabilizes CTR phase retrieval by enforcing the measured spectral amplitude as a hard constraint while allowing additional differentiable beamline components to be attached without requiring explicit inverse propagators.
By providing a bridge between classical phase retrieval and modern differentiable programming, this approach enables the seamless integration of physics-based bunch diagnostics into larger machine-learning optimization and control loops.

\ack{We acknowledge support from the Helmholtz Association via ZT-I-PF-4-027 (SmartPhase) and under the Accelerator Research and Development (ARD) topic of the Helmholtz Matter and Technologies program, and (05D23CR1) VIPR (ErUM: Data).
}

% \funding{This project has received funding by the European Union’s HORIZON-INFRA-2022-TECH-01 call under Grant Agreement Number 10 109 5207. THRILL—Technological and Higher-power laser Research Infrastructures Laboratories.}
% This section is a list of funder names and grant numbers

% \roles{Sample text inserted for demonstration.}
% % List author names and the contributions made to the article, using terms from the NISO Contributor Roles Taxonomy (CRediT) https://credit.niso.org

% \data{Sample text inserted for demonstration.}
% % For more information on IOP Publishing's research data policy see: https://publishingsupport.iopscience.iop.org/questions/research-data/

% \suppdata{Sample text inserted for demonstration.}

\printbibliography

\end{document}